\documentclass[aps,amsmath,showpacs,prl,twocolumn]{revtex4-1}
\pdfoutput=1

\usepackage{amsmath,amssymb}    
\usepackage{graphicx}   
\usepackage{verbatim}   
\usepackage{color}      
\usepackage{hyperref}   
\begin{document}

\title{Generation of large-scale winds in horizontally anisotropic convection}
\author{J.\ von Hardenberg$^1$, D.\ Goluskin$^{2,3}$, A.\ Provenzale$^4$, E.\ A.\ Spiegel$^5$}

\affiliation{$^1$ ISAC-CNR, Torino, Italy; 
$^2$ Mathematics Department, University of Michigan, Ann Arbor MI, USA; \\
$^3$ Center for the Study of Complex Systems, University of Michigan, Ann Arbor MI, USA;\\  $^4$ IGG-CNR, Pisa, Italy;  $^5$ Astronomy Department, Columbia University, New York NY, USA}

\begin{abstract}
\noindent We simulate three-dimensional, horizontally periodic Rayleigh-B\'enard convection, confined between free-slip horizontal plates and rotating about a distant horizontal axis. When both the temperature difference between the plates and the rotation rate are sufficiently large, a strong horizontal wind is generated that is perpendicular to both the rotation vector and the gravity vector. The wind is turbulent, large-scale, and vertically sheared. Horizontal anisotropy, engendered here by rotation, appears necessary for such wind generation. Most of the kinetic energy of the flow resides in the wind, and the vertical turbulent heat flux is much lower on average than when there is no wind.
\end{abstract}
 
\maketitle

Buoyancy-driven convection is a primary mechanism of heat transport in the atmospheres of planets and cool stars, the cores of massive stars, and the Earth's oceans, mantle, and outer core. The essential physics of the process have most commonly been studied in a layer of fluid heated below and cooled above, a configuration called Rayleigh-B\'enard convection (RBC) by \textcite{Avsec1939}. The many natural occurrences of convection and the inherent interest of RBC as a complex system have motivated a great many laboratory experiments and numerical simulations throughout the past century \cite{Siggia1994, Ahlers2009}.

Large-scale horizontal winds in RBC have often been observed, as in the laboratory studies of \textcite{Malkus1954} and \textcite{Krishnamurti1981}. The strongest winds yet reported occur in simulations of horizontally periodic, two-dimensional (2D) RBC between free-slip top and bottom boundaries. These winds have a horizontal wavenumber of zero, and they are vertically sheared, meaning that their velocities change with height. For weak thermal forcing, the bifurcations by which ordinary 2D convective cells can lose stability to states with wind have been studied using the full governing equations \cite{Rucklidge1996, Goluskin2013} and reduced models \cite{Howard1986, Hughes1990, Hermiz1995, Horton1996, Paparella1999, Berning2000, Fitzgerald2014}. When the thermal forcing is strengthened, the wind in 2D simulations intensifies to dominate the underlying convection, and heat transport decreases dramatically \cite{Garcia2003, Goluskin2014}. Comparably strong winds in 3D RBC have not been reported as yet.

The simulations described here reveal that strong, wavenumber-zero winds can arise also in 3D RBC if the horizontal isotropy is broken by uniform rotation about a horizontal axis. The resulting flows, which we call windy convection, are more complicated than their 2D counterparts but can develop winds of comparable strength. Three-dimensional windy convection resembles flow in the equatorial regions of rotating spherical shells (\cite{Christensen2002, Heimpel2007, Kaspi2009, Liao2012}) and in the outer edge of tokamak plasmas in the high-confinement mode \cite{Garcia2006}. Zonal winds in spherical shells are further driven by the way in which rotation and curvature conspire to stretch vortices \cite{Busse1994}, 
but here we find  strong winds  also without this mechanism.

Horizontal anisotropy is apparently necessary for strong wavenumber-zero winds to persist. The winds ceased in our simulations whenever rotation was slowed sufficiently, and winds in the initial conditions invariably died out in 
non-rotating simulations \cite{Parodi2004}. 
Likewise, solutions with mean winds have been found in truncated modal expansions of 3D RBC only when the imposed convective pattern is sufficiently anisotropic \cite{Massaguer1992}. Under horizontally isotropic conditions, other sorts of large-scale circulation have been seen, but none has had much effect on the mean heat flux.  Circulation with a horizontal scale comparable to the layer depth can spatially organize smaller-scale plumes but does not inhibit their heat transport \cite{Hardenberg2008, Bailon2010}, and the wind seen by \textcite{Krishnamurti1981} was apparently not strong enough to prevent thermal plumes from traversing the layer.

We adopt the Boussinesq approximation \cite{Rayleigh1916, Chandrasekhar1961}, in which the fluid has constant kinematic viscosity, $\nu$, thermal diffusivity, $\varkappa$, and coefficient of thermal expansion, $\alpha$. The natural units we use are the layer thickness, $d$, the imposed vertical temperature difference between the bounding plates, $\Delta$, and the thermal time, $\tau_{th}=d^2/\varkappa$. Other time scales include the viscous time, $\tau_{vis}=d^2/\nu$, the dynamical time, $\tau_{dyn}=\sqrt{d/(g\alpha\Delta)}$, where $g$ is gravitational acceleration, an effective dissipative time, $\tau_{diss}=\sqrt{\tau_{th}\tau_{vis}}$, and the rotation period.

The ability of buoyancy forces to overcome viscous drag is measured by the Rayleigh number, $R=\left(\tau_{diss}/\tau_{dyn}\right)^2$, and the relative importance of the two dissipative mechanisms is measured by the Prandtl number, $\sigma=\tau_{th}/\tau_{vis}$. 
We set $\sigma=0.71$ (air).
The dimensionless vertical extent is $0\le z\le1$, and we set horizontal periods of $2\pi$ in the $x$ and $y$ directions. The system rotates about a distant axis parallel to the $y$-axis with dimensionless angular speed $\Omega$, expressed in units of $1/\tau_{th}$. The effect of rotation is approximated as spatially uniform.

\begin{figure*}[htb!]
\centering
\includegraphics[width=1\textwidth]{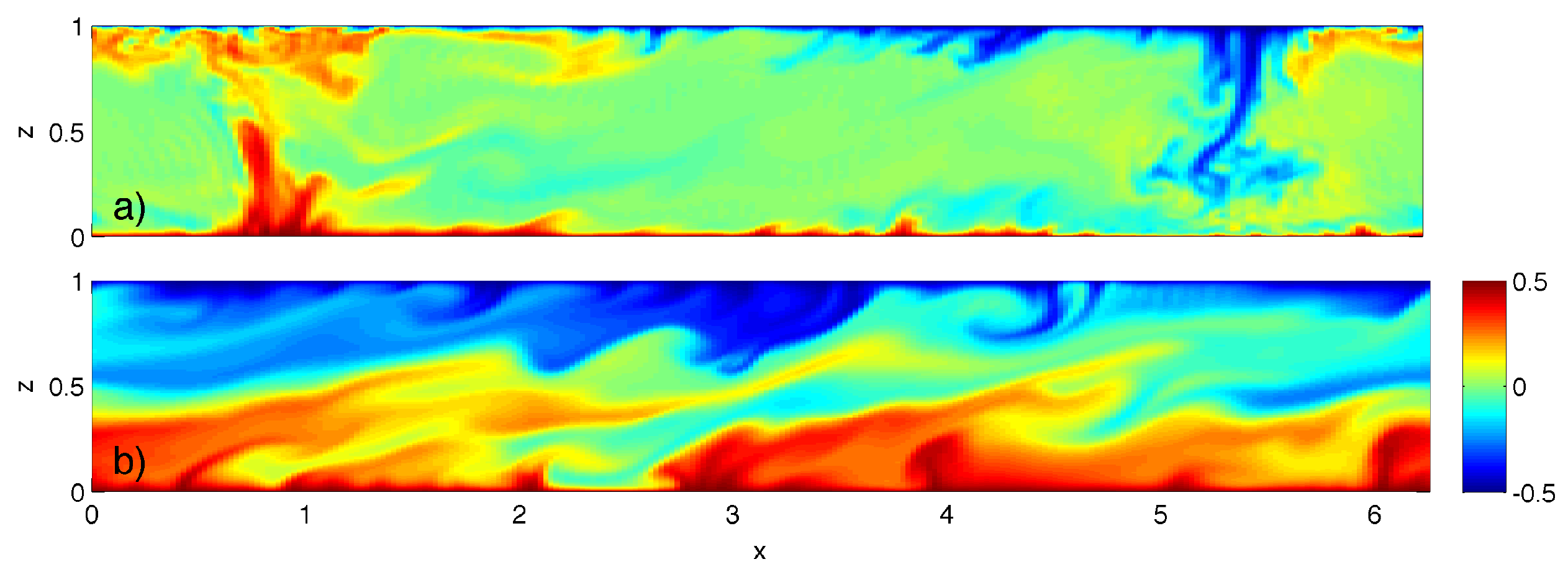}
\caption{\label{fig1} Instantaneous temperature fields in a vertical slice in the 
$(x,z)$ plane at (a) $t=0.04$ and (b) $t=0.25$. The parameters are $R=10^7$ and $2\Omega=10^4$, the top and bottom are free-slip, and the rotation vector points into the page. Slices at other values of $y$ look qualitatively similar since the rotation makes variation in $y$ weak.}
\end{figure*} 

The dimensionless Boussinesq equations governing the velocity $\mathbf{u}=(u,v,w)$, temperature $T$, and pressure $p$ are
\begin{align}
\nabla \cdot \mathbf u &= 0 \label{eq: inc} \\
\partial_t \mathbf u + \mathbf u \cdot \nabla \mathbf u &=
	-\nabla p  + \sigma\nabla^2 \mathbf u + \sigma R\,T\,\hat{\mathbf z}
	-{ {2\Omega}}\,{\hat {\mathbf y}} \times {\mathbf u}  \label{eq: u}  \\
\partial_t T + \mathbf u \cdot \nabla T &= \nabla^2 T. \label{eq: T}
\end{align}
At the top and bottom boundaries, the temperatures are fixed and uniform, 
with free-slip velocity conditions:
\begin{gather}
T|_{z=0}=1/2, \quad T|_{z=1} = -1/2 \\
\partial u/\partial z|_{z=0,1},~\partial v/\partial z|_{z=0,1},~w|_{z=0,1} = 0
\end{gather}
Free-slip boundaries favor strong winds, letting them develop at $R$ that are small enough to be computationally accessible. In 2D, winds arise readily even when only one boundary is free-slip \cite{vanDerPoel2014}.

The governing equations were integrated with a spectral method in the periodic directions with 2/3 dealiasing, second-order finite differences in the vertical, and a third-order fractional step method 
in time
\cite{Parodi2004, Passoni2002}. We used 256 modes in each horizontal direction and 
192 vertical levels,
spaced more closely near the boundaries.

We focus first on a simulation with $R=10^7$ and $2\Omega=10^4$ in which a cyclonic wind develops. Cyclonic winds, with net vorticity parallel to the rotation vector, appear clockwise when observed from the direction of negative~$y$. (Anticyclonic winds are also possible and are discussed below, along with other rotation rates.) Figure 1 shows temperature slices in the $(x,z)$ plane at two instants of the simulation. A movie of this simulation is available at \cite{video}. Beginning from statistically homogeneous, random initial conditions, a pair of turbulent convective cells develop, and the temperature field is dominated by a hot plume and a cold plume (panel a).  This stage persists for many turnover times, during which the cyclonic cell slowly (and non-monotonically) widens while the anticyclonic cell narrows. Eventually, the narrowing cell shrinks out of existence as the hot and cold plumes collide, and the widening cell breaks through to form a cyclonic wind that then dominates the system (panel b). Once the wind forms, fluid flows quickly in the $+x$-direction along the top boundary and in the $-x$-direction along the bottom one. Thermal plumes come into and out of existence, but they are sheared out horizontally and rarely reach the opposite boundaries.

The onset of the wind significantly alters not only the appearance of the flow but also its integral quantities. The specific kinetic energy of motion in the wind direction, $\tfrac{1}{2}\langle u^2\rangle$, is much larger than that in the other two directions, $\tfrac{1}{2}\langle v^2+w^2\rangle$. (Angular brackets denote volume averages.) The decrease in vertical heat flux that accompanies the wind's onset is captured by the Nusselt number, $N(t)=1+\langle wT\rangle$, which is the factor by which convection amplifies heat transport.

\begin{figure*}[htb!]
\centering
\includegraphics[width=0.48\textwidth]{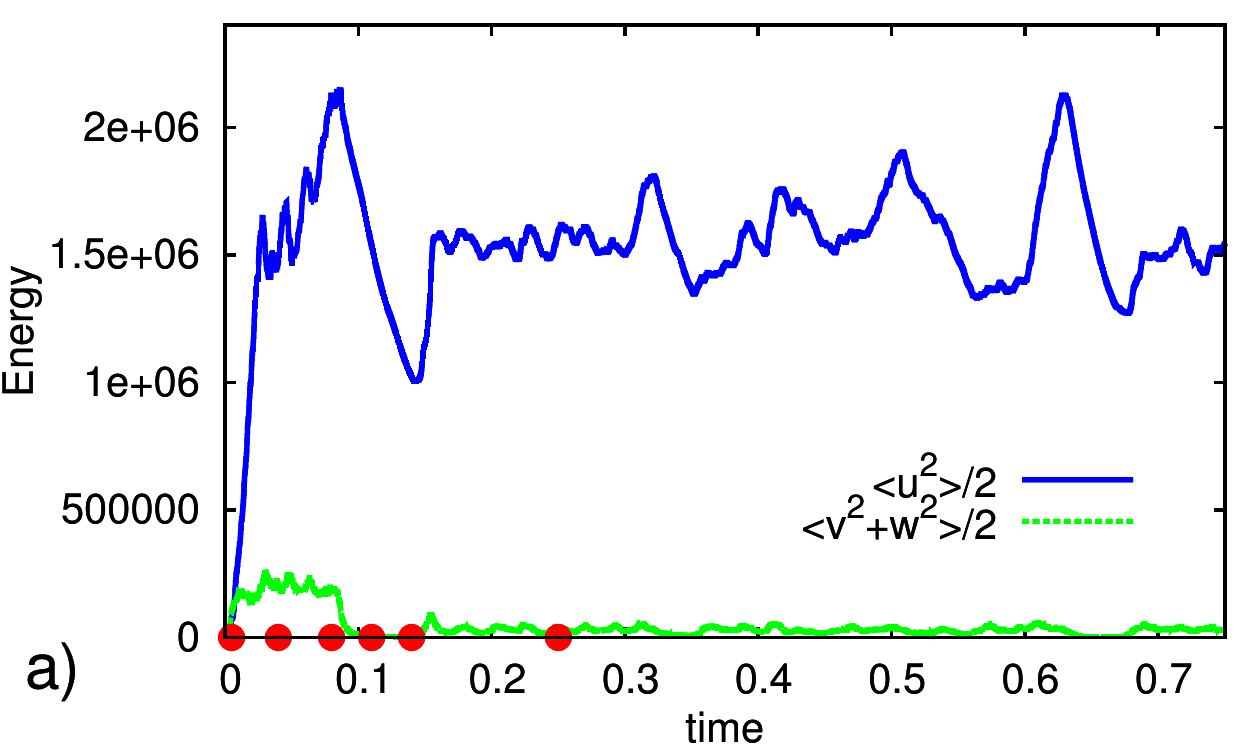}
\hfill
\includegraphics[width=0.48\textwidth]{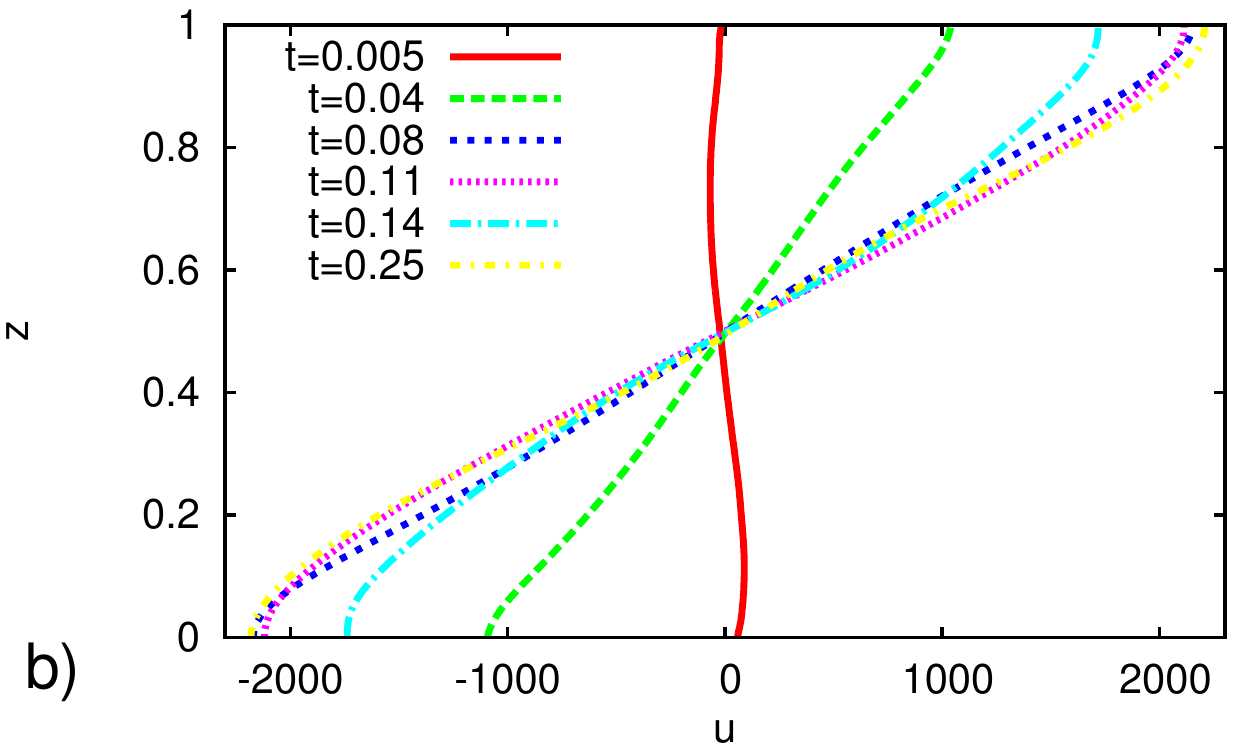}\\
\includegraphics[width=0.48\textwidth]{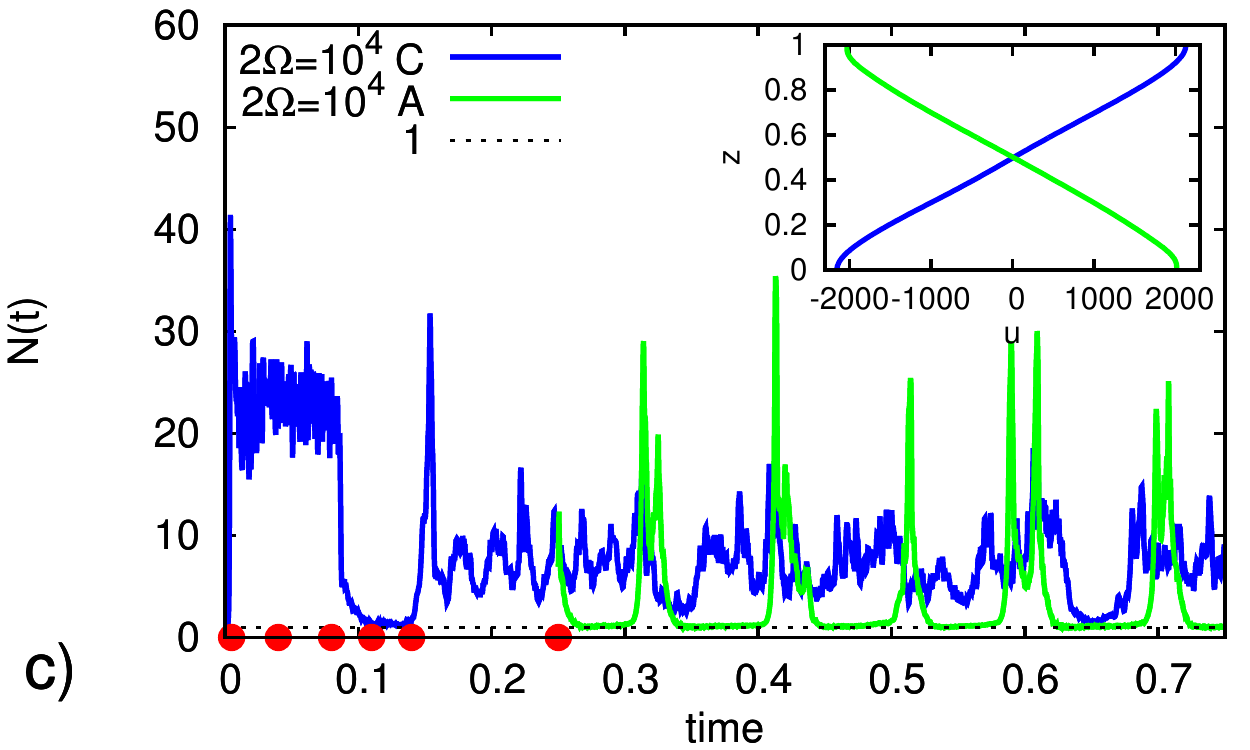}
\hfill
\includegraphics[width=0.48\textwidth]{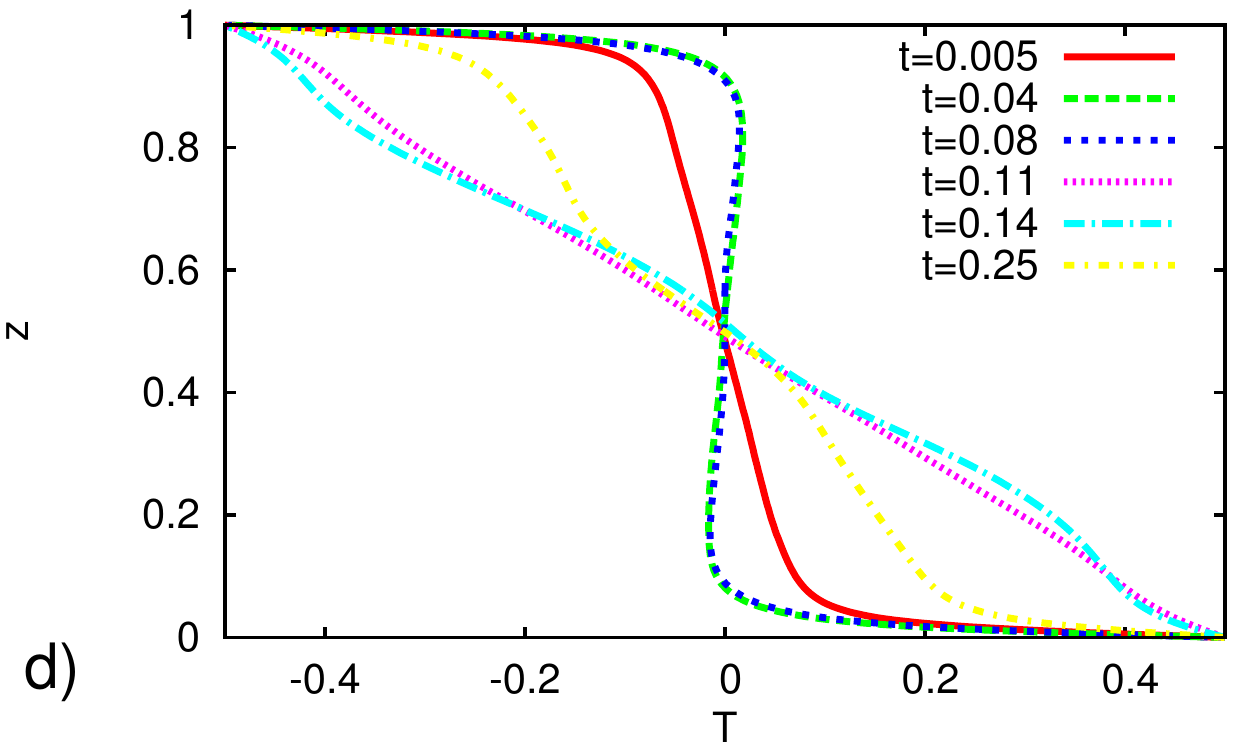}
\caption{\label{fig2}
(a) Time series of volume-averaged specific kinetic energy in the direction of the wind, $\tfrac{1}{2}\langle u^2\rangle$, and normal to the wind, $\tfrac{1}{2}\langle v^2+w^2\rangle$; (b) vertical profiles of the wind velocity averaged horizontally and over 0.025 thermal times; (c) time series of the Nusselt numbers. An anticyclonic simulation (A) is started at time $t=0.25$ by applying the reflection $x\mapsto-x$ to the cyclonic simulation (C). The inset shows mean wind profiles, averaged horizontally and over $0.3\le t\le0.75$;  (d) vertical profiles of the temperature averaged horizontally and over 0.025 thermal times. The dots in panels (a) and (c) mark the starts of the time windows over which the profiles in panels (b) and (d) have been averaged.}
\end{figure*}

Figure 2a shows the time variation of the specific kinetic energy in different velocity components. 
 In an initial transient stage the mean flow slowly strengthens. 
The plumes collide near $t=0.08$, after which the fully developed wind strongly suppresses motions normal to it. Still, these motions never cease completely, as shown by the small but finite values of $\tfrac{1}{2}\langle v^2+w^2\rangle$ in Fig.\ 2a. The growth and saturation of the wind are illustrated in Fig.\ 2b by vertical profiles of $u$, averaged over the horizontal directions and 0.025 thermal times. Even during the transient stage, the wind becomes fairly strong as the cyclonic convective cell grows to dominate the anticyclonic one (see Fig.\ 1a). Ultimately, the vertical profile of $u$ becomes almost linear, with transitory deviations created by plume outbursts.

Figure 2c shows the time evolution of the Nusselt number for the main experiment (blue line), which first displays a strong peak as convective plumes emerge explosively from the linear instability. Then $N(t)$ changes little while the wind builds in strength. When the wind takes over as the cyclonic cell annihilates the anticyclonic one, convective heat transport becomes very feeble, as reflected by $N(t)$ dropping almost to unity. Heat transport recovers only partially thereafter since the system remains dominated by the horizontal wind. The time-averaged Nusselt number is 7.1 after the wind has fully formed ($0.20\le t \le 0.75$), as compared with 22.6 during the transient stage ($0.05\le t\le0.08$). Figure 2d shows the evolution of mean vertical temperature profiles, which are roughly isothermal in the interior before the wind develops but have a strongly unstable stratification afterward.

In addition to being smaller when the wind dominates, $N(t)$ varies more strongly; at irregular intervals, there are deep minima of convective transport during which the Nusselt number dips nearly to its conductive value of unity. These minima occur when the wind becomes especially strong and nearly shuts off the convection from which it draws its energy. Lacking an energy source, the wind decays until it is too weak to inhibit the small-scale convection, which then rebounds. Such behavior is an instance of on-off intermittency \cite{Platt1993}. Very strong intermittency occurs at similar parameters in 2D RBC \cite{Garcia2003, Goluskin2014}, which may be thought of as an extreme limit of horizontal anisotropy in 3D RBC.
 
\begin{figure*}[htb!]
\centering
\includegraphics[width=0.49\textwidth]{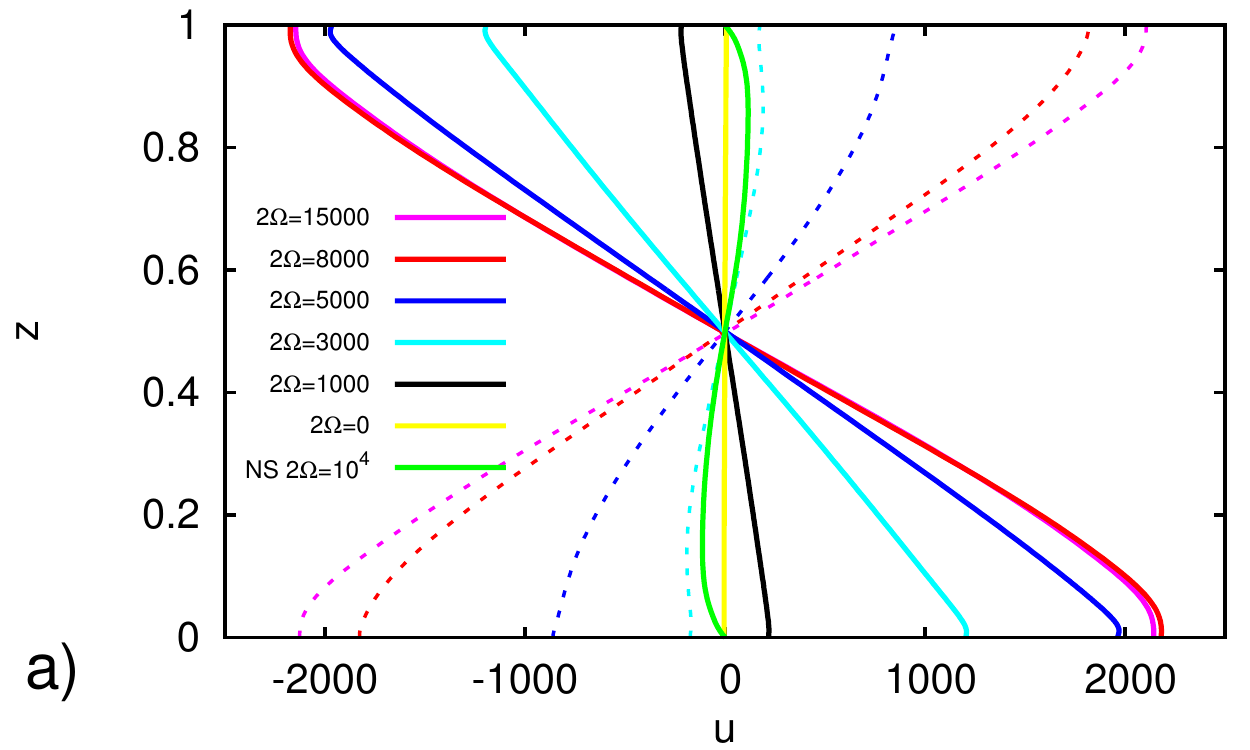}
\hfill
\includegraphics[width=0.49\textwidth]{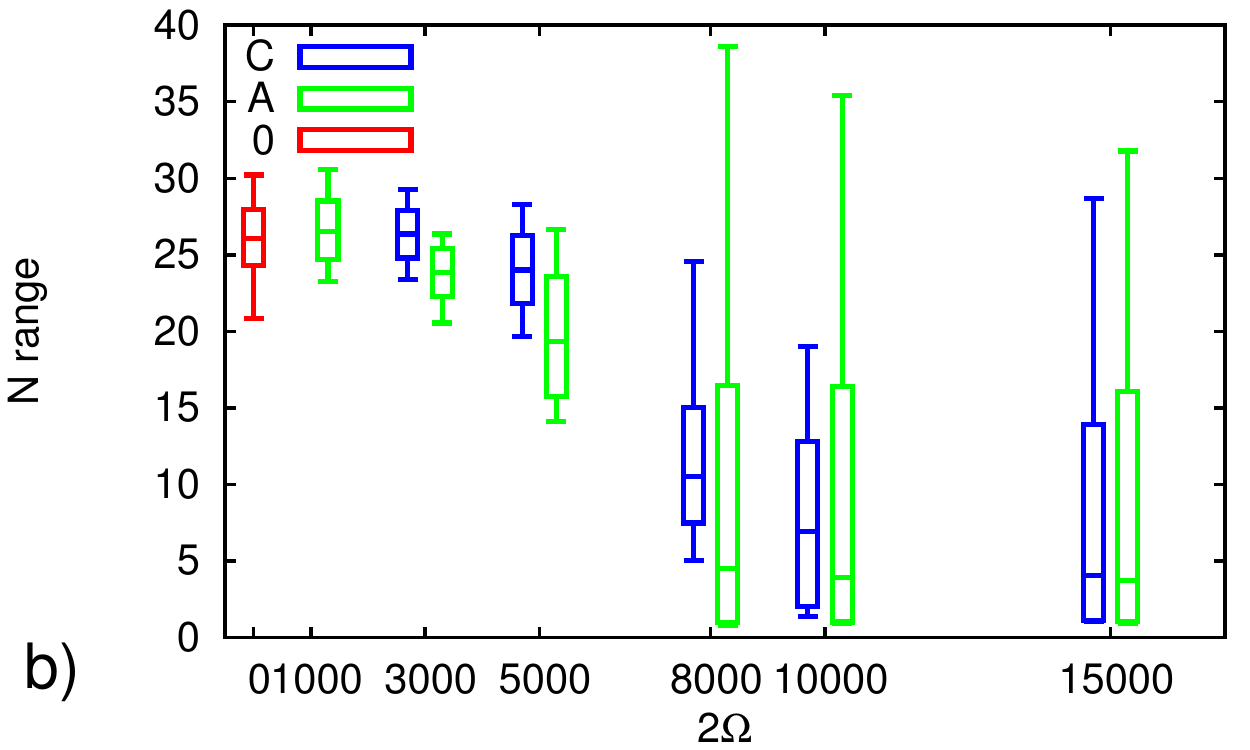}
\caption{\label{fig3}
(a) vertical profiles of the wind velocity, averaged horizontally and over 0.4 thermal times; (b) ranges of $N(t)$ (mean, 5\% and 95\% percentiles, extreme values) explored by the experiments over 0.4 thermal times. Results for rotation rates of $2\Omega = 15000, 8000, 5000, 3000, 1000, 0$ are shown for free-slip boundaries, along with results for $2\Omega = 10^4$ with no-slip (NS) boundaries. Cyclonic (dashed lines) and anticyclonic (continuous lines) windy solutions result from different initial conditions.}
\end{figure*}

In addition to the cyclonic wind shown in Fig.\ 1b and discussed above, there exists at the same parameters a second state with sustained \emph{anti}cyclonic wind---that is, with $u<0$ near the top and $u>0$ near the bottom. We found this anticyclonic state by restarting the cyclonic simulation at $t=0.25$, after reflecting $x\mapsto -x$. Wind profiles and $N(t)$ series for the anticyclonic simulation accompany those for the cyclonic one in Fig.\ 2c. It seems there is bistability between the cyclonic and anticyclonic attractors, each having its own basin of attraction. Fig.\ 2c illustrates that both types of wind are of comparable energy, but the anticyclonic wind bursts much more strongly, resembling bursting seen in 2D \cite{Garcia2003, Goluskin2014, vanDerPoel2014}.

It is rotation that distinguishes between the two wind directions. Without rotation, winds in 2D RBC spontaneously break the reflectional symmetry $x\mapsto -x$ to form a vertical shear with one sign or the other. Here, Coriolis force breaks symmetry not only between the $x$- and $y$-directions but also between cyclonic and anticyclonic winds. For instance, Coriolis forces act upward on fluid flowing in the $+x$ direction and downward on fluid flowing in the $-x$ direction. (The broken $x\mapsto -x$ symmetry recalls the cyclone-anticyclone asymmetry seen in stably stratified geophysical turbulence \citep{Roullet2010}.) Coriolis force cannot help sustain the wind against viscous dissipation, however, since it performs no work on the fluid. With or without rotation, the wind can be driven only by the transfer of energy, via the velocity nonlinearity, from the smaller-scale motions that are buoyantly driven.

Keeping $R=10^7$ and free-slip boundaries, we have explored the effects of rotation rate. Figure 3a gives mean horizontal velocity profiles for both cyclonic and anticyclonic states at various rotation rates. Bistability between each type of wind was found in all cases with $2\Omega\ge3000$. (The windy state of Fig.\ 1b was used to initialize the cyclonic simulations, and its $x\mapsto -x$ reflection was used to initialize most anticyclonic ones.) In both the cyclonic and anticyclonic states, the mean winds grow as rotation is strengthened. Fully developed windy convection like that in Fig.\ 1b, where the wind shear is strong enough to inhibit convection rolls, occurs only for sufficiently high rotation rates. In our simulations, true windy convection occurs only when $2\Omega\ge5000$ in the anticyclonic cases and when $2\Omega\ge8000$ in the cyclonic ones. At lower rotation rates we always find a pair of convective rolls, coexisting with a mean wind that is too weak to destroy them. Fig. 3a also reports a simulation with no-slip boundary conditions. A slight wind develops in this case, but it is not nearly strong enough to destroy the convective cells and create 
true windy convection. 

Figure 3b summarizes the ranges of $N(t)$ found in simulations at various $2\Omega$. At the lower rotation rates where rolls persist, $N(t)$ behaves much as it does in the non-rotating case, where there is no mean wind. 
When the rotation rate is high enough for the wind to preclude convective overturning, we find the dramatically altered heat transport described above for our main simulation: $N(t)$ is highly intermittent, and its time average is small. It seems that the wind is strong enough to significantly alter $N(t)$ only when it is strong enough to prohibit rolls---that is, when $u$ has a single sign near the top boundary and the opposite sign near the bottom one.

In summary, we have observed the formation of strong horizontal winds in 3D RBC when a forced symmetry breaking, here brought about by imposed rotation, is strong enough. When rotation is too weak for wind to be sustained, any wind that starts to form is quickly disrupted by motions normal to it. This is natural because the wind does not suppress convective instability normal to it, whereas the wind does partially suppress motions that are not normal to it and from which it draws its energy. This effect has been demonstrated in the related system where Couette shear is imposed on RBC \cite{Deardorff1965, Zaleski1991}. The tendency of strong winds to quench the motions that generate them has analogs also in the large-scale magnetic fields of convective dynamo theory \cite{Glatzmaier1995, Proctor2014} and the zonal flows of tokamak plasmas \cite{Diamond2005}.

Our findings support the conjecture that the upscale cascade of energy in anisotropic turbulent convection, which here drives sheared winds, drives differential rotation in the equatorial regions of planetary atmospheres and stellar convective zones. Such differential rotation may have interesting consequences. At the top and bottom of the solar convection zone, for example, shear flows have been detected. The shearing flow at the bottom -- the tachocline \cite{Hughes2007} -- is suspected of generating the solar activity that is manifested by sunspots \cite{Spiegel2007}, and the intermittent behavior of the heat transport described here recalls the irregularity of such activity \cite{Spiegel1980}. Other possible locations of windy convection are in the convective cores of massive stars that rotate rapidly. Large-scale shear arising there might influence nuclear reaction rates \cite{Spiegel1984}, with implications for stellar evolution. More immediately, the desire to extrapolate windy convection to astrophysically or geophysically interesting regimes calls for further exploration of parameter space.

\bibliographystyle{apsrev4-1}
\bibliography{rbshear}

\begin{thebibliography}{42}%
\makeatletter
\providecommand \@ifxundefined [1]{%
 \@ifx{#1\undefined}
}%
\providecommand \@ifnum [1]{%
 \ifnum #1\expandafter \@firstoftwo
 \else \expandafter \@secondoftwo
 \fi
}%
\providecommand \@ifx [1]{%
 \ifx #1\expandafter \@firstoftwo
 \else \expandafter \@secondoftwo
 \fi
}%
\providecommand \natexlab [1]{#1}%
\providecommand \enquote  [1]{``#1''}%
\providecommand \bibnamefont  [1]{#1}%
\providecommand \bibfnamefont [1]{#1}%
\providecommand \citenamefont [1]{#1}%
\providecommand \href@noop [0]{\@secondoftwo}%
\providecommand \href [0]{\begingroup \@sanitize@url \@href}%
\providecommand \@href[1]{\@@startlink{#1}\@@href}%
\providecommand \@@href[1]{\endgroup#1\@@endlink}%
\providecommand \@sanitize@url [0]{\catcode `\\12\catcode `\$12\catcode
  `\&12\catcode `\#12\catcode `\^12\catcode `\_12\catcode `\%12\relax}%
\providecommand \@@startlink[1]{}%
\providecommand \@@endlink[0]{}%
\providecommand \url  [0]{\begingroup\@sanitize@url \@url }%
\providecommand \@url [1]{\endgroup\@href {#1}{\urlprefix }}%
\providecommand \urlprefix  [0]{URL }%
\providecommand \Eprint [0]{\href }%
\@ifxundefined \urlstyle {%
  \providecommand \doi  [0]{\begingroup \@sanitize@url \@doi}%
  \providecommand \@doi [1]{\endgroup \@@startlink {\doibase
  #1}doi:\discretionary {}{}{}#1\@@endlink }%
}{%
  \providecommand \doi  [0]{doi:\discretionary{}{}{}\begingroup
  \urlstyle{rm}\Url }%
}%
\providecommand \doibase [0]{http://dx.doi.org/}%
\providecommand \Doi [0]{\begingroup \@sanitize@url \@Doi }%
\providecommand \@Doi  [1]{\endgroup\@@startlink{\doibase#1}\@@Doi}%
\providecommand \@@Doi [1]{#1\@@endlink}%
\providecommand \selectlanguage [0]{\@gobble}%
\providecommand \bibinfo  [0]{\@secondoftwo}%
\providecommand \bibfield  [0]{\@secondoftwo}%
\providecommand \translation [1]{[#1]}%
\providecommand \BibitemOpen [0]{}%
\providecommand \bibitemStop [0]{}%
\providecommand \bibitemNoStop [0]{.\EOS\space}%
\providecommand \EOS [0]{\spacefactor3000\relax}%
\providecommand \BibitemShut  [1]{\csname bibitem#1\endcsname}%
\bibitem [{\citenamefont {Avsec}(1939)}]{Avsec1939}%
  \BibitemOpen
  \bibfield  {author} {\bibinfo {author} {\bibfnamefont {D.}~\bibnamefont
  {Avsec}},\ }\emph {\bibinfo {title} {Tourbillons Thermoconvectifs Dans l'Air.
  Application \`a la m\'et\'eorologie}},\ \href@noop {} {Ph.D. thesis},\
  \bibinfo  {school} {Universit\'e de Paris} (\bibinfo {year}
  {1939})\BibitemShut {NoStop}%
\bibitem [{\citenamefont {Siggia}(1994)}]{Siggia1994}%
  \BibitemOpen
  \bibfield  {author} {\bibinfo {author} {\bibfnamefont {E.~D.}\ \bibnamefont
  {Siggia}},\ }\href@noop {} {\bibfield  {journal} {\bibinfo  {journal} {Annual
  review of fluid mechanics},\ }\textbf {\bibinfo {volume} {26}},\ \bibinfo
  {pages} {137} (\bibinfo {year} {1994})}\BibitemShut {NoStop}%
\bibitem [{\citenamefont {Ahlers}\ \emph {et~al.}(2009)\citenamefont {Ahlers},
  \citenamefont {Grossmann},\ and\ \citenamefont {Lohse}}]{Ahlers2009}%
  \BibitemOpen
  \bibfield  {author} {\bibinfo {author} {\bibfnamefont {G.}~\bibnamefont
  {Ahlers}}, \bibinfo {author} {\bibfnamefont {S.}~\bibnamefont {Grossmann}}, \
  and\ \bibinfo {author} {\bibfnamefont {D.}~\bibnamefont {Lohse}},\
  }\href@noop {} {\bibfield  {journal} {\bibinfo  {journal} {Reviews of Modern
  Physics},\ }\textbf {\bibinfo {volume} {81}},\ \bibinfo {pages} {503}
  (\bibinfo {year} {2009})}\BibitemShut {NoStop}%
\bibitem [{\citenamefont {Malkus}(1954)}]{Malkus1954}%
  \BibitemOpen
  \bibfield  {author} {\bibinfo {author} {\bibfnamefont {W.~V.~R.}\
  \bibnamefont {Malkus}},\ }\href@noop {} {\bibfield  {journal} {\bibinfo
  {journal} {Proceedings of the Royal Society of London. Series A. Mathematical
  and Physical Sciences},\ }\textbf {\bibinfo {volume} {225}},\ \bibinfo
  {pages} {196} (\bibinfo {year} {1954})}\BibitemShut {NoStop}%
\bibitem [{\citenamefont {Krishnamurti}\ and\ \citenamefont
  {Howard}(1981)}]{Krishnamurti1981}%
  \BibitemOpen
  \bibfield  {author} {\bibinfo {author} {\bibfnamefont {R.}~\bibnamefont
  {Krishnamurti}}\ and\ \bibinfo {author} {\bibfnamefont {L.~N.}\ \bibnamefont
  {Howard}},\ }\href@noop {} {\bibfield  {journal} {\bibinfo  {journal}
  {Proceedings of the National Academy of Sciences},\ }\textbf {\bibinfo
  {volume} {78}} (\bibinfo {year} {1981})}\BibitemShut {NoStop}%
\bibitem [{\citenamefont {Rucklidge}\ and\ \citenamefont
  {Matthews}(1996)}]{Rucklidge1996}%
  \BibitemOpen
  \bibfield  {author} {\bibinfo {author} {\bibfnamefont {A.~M.}\ \bibnamefont
  {Rucklidge}}\ and\ \bibinfo {author} {\bibfnamefont {P.~C.}\ \bibnamefont
  {Matthews}},\ }\href@noop {} {\bibfield  {journal} {\bibinfo  {journal}
  {Nonlinearity},\ }\textbf {\bibinfo {volume} {9}},\ \bibinfo {pages} {311}
  (\bibinfo {year} {1996})}\BibitemShut {NoStop}%
\bibitem [{\citenamefont {Goluskin}(2013)}]{Goluskin2013}%
  \BibitemOpen
  \bibfield  {author} {\bibinfo {author} {\bibfnamefont {D.}~\bibnamefont
  {Goluskin}},\ }\emph {\bibinfo {title} {Zonal flow driven by convection and
  convection driven by internal heating}},\ \href@noop {} {Ph.D. thesis},\
  \bibinfo  {school} {Graduate School of Arts and Sciences, Columbia
  University} (\bibinfo {year} {2013})\BibitemShut {NoStop}%
\bibitem [{\citenamefont {Howard}\ and\ \citenamefont
  {Krishnamurti}(1986)}]{Howard1986}%
  \BibitemOpen
  \bibfield  {author} {\bibinfo {author} {\bibfnamefont {L.~N.}\ \bibnamefont
  {Howard}}\ and\ \bibinfo {author} {\bibfnamefont {R.}~\bibnamefont
  {Krishnamurti}},\ }\href@noop {} {\bibfield  {journal} {\bibinfo  {journal}
  {Journal of Fluid Mechanics},\ }\textbf {\bibinfo {volume} {170}},\ \bibinfo
  {pages} {385} (\bibinfo {year} {1986})}\BibitemShut {NoStop}%
\bibitem [{\citenamefont {Hughes}\ and\ \citenamefont
  {Proctor}(1990)}]{Hughes1990}%
  \BibitemOpen
  \bibfield  {author} {\bibinfo {author} {\bibfnamefont {D.~W.}\ \bibnamefont
  {Hughes}}\ and\ \bibinfo {author} {\bibfnamefont {M.~R.~E.}\ \bibnamefont
  {Proctor}},\ }\href@noop {} {\bibfield  {journal} {\bibinfo  {journal}
  {Nonlinearity},\ }\textbf {\bibinfo {volume} {3}},\ \bibinfo {pages} {127}
  (\bibinfo {year} {1990})}\BibitemShut {NoStop}%
\bibitem [{\citenamefont {Hermiz}\ \emph {et~al.}(1995)\citenamefont {Hermiz},
  \citenamefont {Guzdar},\ and\ \citenamefont {Finn}}]{Hermiz1995}%
  \BibitemOpen
  \bibfield  {author} {\bibinfo {author} {\bibfnamefont {K.~B.}\ \bibnamefont
  {Hermiz}}, \bibinfo {author} {\bibfnamefont {P.~N.}\ \bibnamefont {Guzdar}},
  \ and\ \bibinfo {author} {\bibfnamefont {J.~M.}\ \bibnamefont {Finn}},\
  }\href@noop {} {\bibfield  {journal} {\bibinfo  {journal} {Physical Review
  E},\ }\textbf {\bibinfo {volume} {51}},\ \bibinfo {pages} {325} (\bibinfo
  {year} {1995})}\BibitemShut {NoStop}%
\bibitem [{\citenamefont {Horton}\ \emph {et~al.}(1996)\citenamefont {Horton},
  \citenamefont {Hu},\ and\ \citenamefont {Laval}}]{Horton1996}%
  \BibitemOpen
  \bibfield  {author} {\bibinfo {author} {\bibfnamefont {W.}~\bibnamefont
  {Horton}}, \bibinfo {author} {\bibfnamefont {G.}~\bibnamefont {Hu}}, \ and\
  \bibinfo {author} {\bibfnamefont {G.}~\bibnamefont {Laval}},\ }\href@noop {}
  {\bibfield  {journal} {\bibinfo  {journal} {Physics of Plasmas},\ }\textbf
  {\bibinfo {volume} {3}},\ \bibinfo {pages} {2912} (\bibinfo {year}
  {1996})}\BibitemShut {NoStop}%
\bibitem [{\citenamefont {Paparella}\ and\ \citenamefont
  {Spiegel}(1999)}]{Paparella1999}%
  \BibitemOpen
  \bibfield  {author} {\bibinfo {author} {\bibfnamefont {F.}~\bibnamefont
  {Paparella}}\ and\ \bibinfo {author} {\bibfnamefont {E.~A.}\ \bibnamefont
  {Spiegel}},\ }\href@noop {} {\bibfield  {journal} {\bibinfo  {journal}
  {Physics of Fluids},\ }\textbf {\bibinfo {volume} {11}},\ \bibinfo {pages}
  {1161} (\bibinfo {year} {1999})}\BibitemShut {NoStop}%
\bibitem [{\citenamefont {Berning}\ and\ \citenamefont
  {Spatschek}(2000)}]{Berning2000}%
  \BibitemOpen
  \bibfield  {author} {\bibinfo {author} {\bibfnamefont {M.}~\bibnamefont
  {Berning}}\ and\ \bibinfo {author} {\bibfnamefont {K.~H.}\ \bibnamefont
  {Spatschek}},\ }\href@noop {} {\bibfield  {journal} {\bibinfo  {journal}
  {Physical Review E},\ }\textbf {\bibinfo {volume} {62}},\ \bibinfo {pages}
  {1162} (\bibinfo {year} {2000})}\BibitemShut {NoStop}%
\bibitem [{\citenamefont {Fitzgerald}\ and\ \citenamefont
  {Farrell}(2014)}]{Fitzgerald2014}%
  \BibitemOpen
  \bibfield  {author} {\bibinfo {author} {\bibfnamefont {J.~G.}\ \bibnamefont
  {Fitzgerald}}\ and\ \bibinfo {author} {\bibfnamefont {B.~F.}\ \bibnamefont
  {Farrell}},\ }\Doi {http://dx.doi.org/10.1063/1.4875814} {\bibfield
  {journal} {\bibinfo  {journal} {Physics of Fluids},\ }\textbf {\bibinfo
  {volume} {26}},\ \bibinfo {eid} {054104} (\bibinfo {year}
  {2014})}\BibitemShut {NoStop}%
\bibitem [{\citenamefont {Garcia}\ \emph {et~al.}(2003)\citenamefont {Garcia},
  \citenamefont {Bian}, \citenamefont {Paulsen}, \citenamefont {Benkadda},\
  and\ \citenamefont {Rypdal}}]{Garcia2003}%
  \BibitemOpen
  \bibfield  {author} {\bibinfo {author} {\bibfnamefont {O.~E.}\ \bibnamefont
  {Garcia}}, \bibinfo {author} {\bibfnamefont {N.~H.}\ \bibnamefont {Bian}},
  \bibinfo {author} {\bibfnamefont {J.~V.}\ \bibnamefont {Paulsen}}, \bibinfo
  {author} {\bibfnamefont {S.}~\bibnamefont {Benkadda}}, \ and\ \bibinfo
  {author} {\bibfnamefont {K.}~\bibnamefont {Rypdal}},\ }\href@noop {}
  {\bibfield  {journal} {\bibinfo  {journal} {Plasma physics and controlled
  fusion},\ }\textbf {\bibinfo {volume} {45}},\ \bibinfo {pages} {919}
  (\bibinfo {year} {2003})}\BibitemShut {NoStop}%
\bibitem [{\citenamefont {Goluskin}\ \emph {et~al.}(2014)\citenamefont
  {Goluskin}, \citenamefont {Johnston}, \citenamefont {Flierl},\ and\
  \citenamefont {Spiegel}}]{Goluskin2014}%
  \BibitemOpen
  \bibfield  {author} {\bibinfo {author} {\bibfnamefont {D.}~\bibnamefont
  {Goluskin}}, \bibinfo {author} {\bibfnamefont {H.}~\bibnamefont {Johnston}},
  \bibinfo {author} {\bibfnamefont {G.~R.}\ \bibnamefont {Flierl}}, \ and\
  \bibinfo {author} {\bibfnamefont {E.~A.}\ \bibnamefont {Spiegel}},\
  }\href@noop {} {\bibfield  {journal} {\bibinfo  {journal} {Journal of Fluid
  Mechanics},\ }\textbf {\bibinfo {volume} {759}},\ \bibinfo {pages} {360}
  (\bibinfo {year} {2014})}\BibitemShut {NoStop}%
\bibitem [{\citenamefont {Christensen}(2002)}]{Christensen2002}%
  \BibitemOpen
  \bibfield  {author} {\bibinfo {author} {\bibfnamefont {U.~R.}\ \bibnamefont
  {Christensen}},\ }\href@noop {} {\bibfield  {journal} {\bibinfo  {journal}
  {Journal of Fluid Mechanics},\ }\textbf {\bibinfo {volume} {470}},\ \bibinfo
  {pages} {115} (\bibinfo {year} {2002})}\BibitemShut {NoStop}%
\bibitem [{\citenamefont {Heimpel}\ and\ \citenamefont
  {Aurnou}(2007)}]{Heimpel2007}%
  \BibitemOpen
  \bibfield  {author} {\bibinfo {author} {\bibfnamefont {M.}~\bibnamefont
  {Heimpel}}\ and\ \bibinfo {author} {\bibfnamefont {J.}~\bibnamefont
  {Aurnou}},\ }\href@noop {} {\bibfield  {journal} {\bibinfo  {journal}
  {Icarus},\ }\textbf {\bibinfo {volume} {187}},\ \bibinfo {pages} {540}
  (\bibinfo {year} {2007})}\BibitemShut {NoStop}%
\bibitem [{\citenamefont {Kaspi}\ \emph {et~al.}(2009)\citenamefont {Kaspi},
  \citenamefont {Flierl},\ and\ \citenamefont {Showman}}]{Kaspi2009}%
  \BibitemOpen
  \bibfield  {author} {\bibinfo {author} {\bibfnamefont {Y.}~\bibnamefont
  {Kaspi}}, \bibinfo {author} {\bibfnamefont {G.~R.}\ \bibnamefont {Flierl}}, \
  and\ \bibinfo {author} {\bibfnamefont {A.~P.}\ \bibnamefont {Showman}},\
  }\href@noop {} {\bibfield  {journal} {\bibinfo  {journal} {Icarus},\ }\textbf
  {\bibinfo {volume} {202}},\ \bibinfo {pages} {525} (\bibinfo {year}
  {2009})}\BibitemShut {NoStop}%
\bibitem [{\citenamefont {Liao}\ \emph {et~al.}(2012)\citenamefont {Liao},
  \citenamefont {Zhang},\ and\ \citenamefont {Li}}]{Liao2012}%
  \BibitemOpen
  \bibfield  {author} {\bibinfo {author} {\bibfnamefont {X.}~\bibnamefont
  {Liao}}, \bibinfo {author} {\bibfnamefont {K.}~\bibnamefont {Zhang}}, \ and\
  \bibinfo {author} {\bibfnamefont {L.}~\bibnamefont {Li}},\ }\href@noop {}
  {\bibfield  {journal} {\bibinfo  {journal} {Geophysical \& Astrophysical
  Fluid Dynamics},\ }\textbf {\bibinfo {volume} {106}},\ \bibinfo {pages} {643}
  (\bibinfo {year} {2012})}\BibitemShut {NoStop}%
\bibitem [{\citenamefont {Garcia}\ \emph {et~al.}(2006)\citenamefont {Garcia},
  \citenamefont {Bian}, \citenamefont {Naulin}, \citenamefont {Nielsen},\ and\
  \citenamefont {Rasmussen}}]{Garcia2006}%
  \BibitemOpen
  \bibfield  {author} {\bibinfo {author} {\bibfnamefont {O.~E.}\ \bibnamefont
  {Garcia}}, \bibinfo {author} {\bibfnamefont {N.~H.}\ \bibnamefont {Bian}},
  \bibinfo {author} {\bibfnamefont {V.}~\bibnamefont {Naulin}}, \bibinfo
  {author} {\bibfnamefont {A.~H.}\ \bibnamefont {Nielsen}}, \ and\ \bibinfo
  {author} {\bibfnamefont {J.~J.}\ \bibnamefont {Rasmussen}},\ }\href@noop {}
  {\bibfield  {journal} {\bibinfo  {journal} {Physica Scripta},\ }\textbf
  {\bibinfo {volume} {2006}},\ \bibinfo {pages} {104} (\bibinfo {year}
  {2006})}\BibitemShut {NoStop}%
\bibitem [{\citenamefont {Busse}(1994)}]{Busse1994}%
  \BibitemOpen
  \bibfield  {author} {\bibinfo {author} {\bibfnamefont {F.~H.}\ \bibnamefont
  {Busse}},\ }\href@noop {} {\bibfield  {journal} {\bibinfo  {journal}
  {Chaos},\ }\textbf {\bibinfo {volume} {4}} (\bibinfo {year}
  {1994})}\BibitemShut {NoStop}%
\bibitem [{\citenamefont {Parodi}\ \emph {et~al.}(2004)\citenamefont {Parodi},
  \citenamefont {von Hardenberg}, \citenamefont {Passoni}, \citenamefont
  {Provenzale},\ and\ \citenamefont {Spiegel}}]{Parodi2004}%
  \BibitemOpen
  \bibfield  {author} {\bibinfo {author} {\bibfnamefont {A.}~\bibnamefont
  {Parodi}}, \bibinfo {author} {\bibfnamefont {J.}~\bibnamefont {von
  Hardenberg}}, \bibinfo {author} {\bibfnamefont {G.}~\bibnamefont {Passoni}},
  \bibinfo {author} {\bibfnamefont {A.}~\bibnamefont {Provenzale}}, \ and\
  \bibinfo {author} {\bibfnamefont {E.~A.}\ \bibnamefont {Spiegel}},\
  }\href@noop {} {\bibfield  {journal} {\bibinfo  {journal} {Physical Review
  Letters},\ }\textbf {\bibinfo {volume} {92}},\ \bibinfo {pages} {194503}
  (\bibinfo {year} {2004})}\BibitemShut {NoStop}%
\bibitem [{\citenamefont {Massaguer}\ \emph {et~al.}(1992)\citenamefont
  {Massaguer}, \citenamefont {Spiegel},\ and\ \citenamefont
  {Zahn}}]{Massaguer1992}%
  \BibitemOpen
  \bibfield  {author} {\bibinfo {author} {\bibfnamefont {J.~M.}\ \bibnamefont
  {Massaguer}}, \bibinfo {author} {\bibfnamefont {E.~A.}\ \bibnamefont
  {Spiegel}}, \ and\ \bibinfo {author} {\bibfnamefont {J.-P.}\ \bibnamefont
  {Zahn}},\ }\href@noop {} {\bibfield  {journal} {\bibinfo  {journal} {Physics
  of Fluids A: Fluid Dynamics (1989-1993)},\ }\textbf {\bibinfo {volume} {4}},\
  \bibinfo {pages} {1333} (\bibinfo {year} {1992})}\BibitemShut {NoStop}%
\bibitem [{\citenamefont {von Hardenberg}\ \emph {et~al.}(2008)\citenamefont
  {von Hardenberg}, \citenamefont {Parodi}, \citenamefont {Passoni},
  \citenamefont {Provenzale},\ and\ \citenamefont {Spiegel}}]{Hardenberg2008}%
  \BibitemOpen
  \bibfield  {author} {\bibinfo {author} {\bibfnamefont {J.}~\bibnamefont {von
  Hardenberg}}, \bibinfo {author} {\bibfnamefont {A.}~\bibnamefont {Parodi}},
  \bibinfo {author} {\bibfnamefont {G.}~\bibnamefont {Passoni}}, \bibinfo
  {author} {\bibfnamefont {A.}~\bibnamefont {Provenzale}}, \ and\ \bibinfo
  {author} {\bibfnamefont {E.~A.}\ \bibnamefont {Spiegel}},\ }\href@noop {}
  {\bibfield  {journal} {\bibinfo  {journal} {Physics Letters A},\ }\textbf
  {\bibinfo {volume} {372}},\ \bibinfo {pages} {2223} (\bibinfo {year}
  {2008})}\BibitemShut {NoStop}%
\bibitem [{\citenamefont {Bailon-Cuba}\ \emph {et~al.}(2010)\citenamefont
  {Bailon-Cuba}, \citenamefont {Emran},\ and\ \citenamefont
  {Schumacher}}]{Bailon2010}%
  \BibitemOpen
  \bibfield  {author} {\bibinfo {author} {\bibfnamefont {J.}~\bibnamefont
  {Bailon-Cuba}}, \bibinfo {author} {\bibfnamefont {M.~S.}\ \bibnamefont
  {Emran}}, \ and\ \bibinfo {author} {\bibfnamefont {J.}~\bibnamefont
  {Schumacher}},\ }\href@noop {} {\bibfield  {journal} {\bibinfo  {journal}
  {Journal of Fluid Mechanics},\ }\textbf {\bibinfo {volume} {655}},\ \bibinfo
  {pages} {152} (\bibinfo {year} {2010})}\BibitemShut {NoStop}%
\bibitem [{\citenamefont {Rayleigh}(1916)}]{Rayleigh1916}%
  \BibitemOpen
  \bibfield  {author} {\bibinfo {author} {\bibfnamefont {J.~W.~S.}\
  \bibnamefont {Rayleigh}},\ }\href@noop {} {\bibfield  {journal} {\bibinfo
  {journal} {The London, Edinburgh, and Dublin Philosophical Magazine and
  Journal of Science},\ }\textbf {\bibinfo {volume} {32}},\ \bibinfo {pages}
  {529} (\bibinfo {year} {1916})}\BibitemShut {NoStop}%
\bibitem [{\citenamefont {Chandrasekhar}(1961)}]{Chandrasekhar1961}%
  \BibitemOpen
  \bibfield  {author} {\bibinfo {author} {\bibfnamefont {S.}~\bibnamefont
  {Chandrasekhar}},\ }\href@noop {} {\emph {\bibinfo {title} {Hydrodynamic and
  hydromagnetic stability}}}\ (\bibinfo  {publisher} {Clarendon Press,
  Oxford},\ \bibinfo {year} {1961})\BibitemShut {NoStop}%
\bibitem [{\citenamefont {van~der Poel}\ \emph {et~al.}(2014)\citenamefont
  {van~der Poel}, \citenamefont {Ostilla-M{\'o}nico}, \citenamefont
  {Verzicco},\ and\ \citenamefont {Lohse}}]{vanDerPoel2014}%
  \BibitemOpen
  \bibfield  {author} {\bibinfo {author} {\bibfnamefont {E.~P.}\ \bibnamefont
  {van~der Poel}}, \bibinfo {author} {\bibfnamefont {R.}~\bibnamefont
  {Ostilla-M{\'o}nico}}, \bibinfo {author} {\bibfnamefont {R.}~\bibnamefont
  {Verzicco}}, \ and\ \bibinfo {author} {\bibfnamefont {D.}~\bibnamefont
  {Lohse}},\ }\href@noop {} {\bibfield  {journal} {\bibinfo  {journal}
  {Physical Review E},\ }\textbf {\bibinfo {volume} {90}},\ \bibinfo {pages}
  {013017} (\bibinfo {year} {2014})}\BibitemShut {NoStop}%
\bibitem [{\citenamefont {Passoni}\ \emph {et~al.}(2002)\citenamefont
  {Passoni}, \citenamefont {Alfonsi},\ and\ \citenamefont
  {Galbiati}}]{Passoni2002}%
  \BibitemOpen
  \bibfield  {author} {\bibinfo {author} {\bibfnamefont {G.}~\bibnamefont
  {Passoni}}, \bibinfo {author} {\bibfnamefont {G.}~\bibnamefont {Alfonsi}}, \
  and\ \bibinfo {author} {\bibfnamefont {M.}~\bibnamefont {Galbiati}},\
  }\href@noop {} {\bibfield  {journal} {\bibinfo  {journal} {Int. J. Numer.
  Meth. Fl.},\ }\textbf {\bibinfo {volume} {38}},\ \bibinfo {pages} {1069}
  (\bibinfo {year} {2002})}\BibitemShut {NoStop}%
\bibitem [{vid()}]{video}%
  \BibitemOpen
  \href@noop {} {}\bibinfo {howpublished}
  {\url{http://www.to.isac.cnr.it/rbshear/movie3d.html}}\BibitemShut {NoStop}%
\bibitem [{\citenamefont {Platt}\ \emph {et~al.}(1993)\citenamefont {Platt},
  \citenamefont {Spiegel},\ and\ \citenamefont {Tresser}}]{Platt1993}%
  \BibitemOpen
  \bibfield  {author} {\bibinfo {author} {\bibfnamefont {N.~S.}\ \bibnamefont
  {Platt}}, \bibinfo {author} {\bibfnamefont {E.~A.}\ \bibnamefont {Spiegel}},
  \ and\ \bibinfo {author} {\bibfnamefont {C.}~\bibnamefont {Tresser}},\
  }\href@noop {} {\bibfield  {journal} {\bibinfo  {journal} {Physical Review
  Letters},\ }\textbf {\bibinfo {volume} {70}},\ \bibinfo {pages} {279}
  (\bibinfo {year} {1993})}\BibitemShut {NoStop}%
\bibitem [{\citenamefont {Roullet}\ and\ \citenamefont
  {Klein}(2010)}]{Roullet2010}%
  \BibitemOpen
  \bibfield  {author} {\bibinfo {author} {\bibfnamefont {G.}~\bibnamefont
  {Roullet}}\ and\ \bibinfo {author} {\bibfnamefont {P.}~\bibnamefont
  {Klein}},\ }\href@noop {} {\bibfield  {journal} {\bibinfo  {journal}
  {Physical Review Letters},\ }\textbf {\bibinfo {volume} {104}},\ \bibinfo
  {pages} {218501} (\bibinfo {year} {2010})}\BibitemShut {NoStop}%
\bibitem [{\citenamefont {Deardorff}(1965)}]{Deardorff1965}%
  \BibitemOpen
  \bibfield  {author} {\bibinfo {author} {\bibfnamefont {J.~W.}\ \bibnamefont
  {Deardorff}},\ }\href@noop {} {\bibfield  {journal} {\bibinfo  {journal}
  {Physics of Fluids (1958-1988)},\ }\textbf {\bibinfo {volume} {8}},\ \bibinfo
  {pages} {1027} (\bibinfo {year} {1965})}\BibitemShut {NoStop}%
\bibitem [{\citenamefont {Zaleski}(1991)}]{Zaleski1991}%
  \BibitemOpen
  \bibfield  {author} {\bibinfo {author} {\bibfnamefont {S.}~\bibnamefont
  {Zaleski}},\ }in\ \Doi {10.1007/978-1-4615-3750-2_15} {\emph {\bibinfo
  {booktitle} {The Global Geometry of Turbulence}}},\ \bibinfo {series} {NATO
  ASI Series}, Vol.\ \bibinfo {volume} {268},\ \bibinfo {editor} {edited by\
  \bibinfo {editor} {\bibfnamefont {J.}~\bibnamefont {Jim\'enez}}}\ (\bibinfo
  {publisher} {Springer US},\ \bibinfo {year} {1991})\ pp.\ \bibinfo {pages}
  {167--179},\ ISBN \bibinfo {isbn} {978-1-4613-6670-6}\BibitemShut {NoStop}%
\bibitem [{\citenamefont {Glatzmaier}\ and\ \citenamefont
  {Roberts}(1995)}]{Glatzmaier1995}%
  \BibitemOpen
  \bibfield  {author} {\bibinfo {author} {\bibfnamefont {G.~A.}\ \bibnamefont
  {Glatzmaier}}\ and\ \bibinfo {author} {\bibfnamefont {P.~H.}\ \bibnamefont
  {Roberts}},\ }\href@noop {} {\bibfield  {journal} {\bibinfo  {journal}
  {Physics of the Earth and Planetary Interiors},\ }\textbf {\bibinfo {volume}
  {91}},\ \bibinfo {pages} {63} (\bibinfo {year} {1995})}\BibitemShut {NoStop}%
\bibitem [{\citenamefont {Weiss}\ and\ \citenamefont
  {Proctor}(2014)}]{Proctor2014}%
  \BibitemOpen
  \bibfield  {author} {\bibinfo {author} {\bibfnamefont {N.~O.}\ \bibnamefont
  {Weiss}}\ and\ \bibinfo {author} {\bibfnamefont {M.~R.~E.}\ \bibnamefont
  {Proctor}},\ }\href@noop {} {\emph {\bibinfo {title} {Magnetoconvection}}}\
  (\bibinfo  {publisher} {Cambridge University Press},\ \bibinfo {year}
  {2014})\BibitemShut {NoStop}%
\bibitem [{\citenamefont {Diamond}\ \emph {et~al.}(2005)\citenamefont
  {Diamond}, \citenamefont {Itoh}, \citenamefont {Itoh},\ and\ \citenamefont
  {Hahm}}]{Diamond2005}%
  \BibitemOpen
  \bibfield  {author} {\bibinfo {author} {\bibfnamefont {P.~H.}\ \bibnamefont
  {Diamond}}, \bibinfo {author} {\bibfnamefont {S.}~\bibnamefont {Itoh}},
  \bibinfo {author} {\bibfnamefont {K.}~\bibnamefont {Itoh}}, \ and\ \bibinfo
  {author} {\bibfnamefont {T.}~\bibnamefont {Hahm}},\ }\href@noop {} {\bibfield
   {journal} {\bibinfo  {journal} {Plasma Physics and Controlled Fusion},\
  }\textbf {\bibinfo {volume} {47}},\ \bibinfo {pages} {R35} (\bibinfo {year}
  {2005})}\BibitemShut {NoStop}%
\bibitem [{\citenamefont {Hughes}\ \emph {et~al.}(2007)\citenamefont {Hughes},
  \citenamefont {Rosner},\ and\ \citenamefont {Weiss}}]{Hughes2007}%
  \BibitemOpen
  \bibfield  {author} {\bibinfo {author} {\bibfnamefont {D.~W.}\ \bibnamefont
  {Hughes}}, \bibinfo {author} {\bibfnamefont {R.}~\bibnamefont {Rosner}}, \
  and\ \bibinfo {author} {\bibfnamefont {N.}~\bibnamefont {Weiss}},\
  }\href@noop {} {\emph {\bibinfo {title} {The Solar Tachocline}}}\ (\bibinfo
  {publisher} {Cambridge University Press},\ \bibinfo {year}
  {2007})\BibitemShut {NoStop}%
\bibitem [{\citenamefont {Spiegel}(2007)}]{Spiegel2007}%
  \BibitemOpen
  \bibfield  {author} {\bibinfo {author} {\bibfnamefont {E.~A.}\ \bibnamefont
  {Spiegel}},\ }\enquote {\bibinfo {title} {Reflections on the solar
  tachocline},}\ in\ \href@noop {} {\emph {\bibinfo {booktitle} {The Solar
  Tachocline}}},\ \bibinfo {editor} {edited by\ \bibinfo {editor}
  {\bibfnamefont {D.~W.}\ \bibnamefont {Hughes}}, \bibinfo {editor}
  {\bibfnamefont {R.}~\bibnamefont {Rosner}}, \ and\ \bibinfo {editor}
  {\bibfnamefont {W.~N.}\ \bibnamefont {O.}}}\ (\bibinfo  {publisher}
  {Cambridge University Press},\ \bibinfo {year} {2007})\BibitemShut {NoStop}%
\bibitem [{\citenamefont {Spiegel}\ and\ \citenamefont
  {Weiss}(1980)}]{Spiegel1980}%
  \BibitemOpen
  \bibfield  {author} {\bibinfo {author} {\bibfnamefont {E.~A.}\ \bibnamefont
  {Spiegel}}\ and\ \bibinfo {author} {\bibfnamefont {N.~O.}\ \bibnamefont
  {Weiss}},\ }\href@noop {} {\bibfield  {journal} {\bibinfo  {journal}
  {Nature},\ \bibinfo {pages} {616}} (\bibinfo {year} {1980})}\BibitemShut
  {NoStop}%
\bibitem [{\citenamefont {Spiegel}\ and\ \citenamefont
  {Zaleski}(1984)}]{Spiegel1984}%
  \BibitemOpen
  \bibfield  {author} {\bibinfo {author} {\bibfnamefont {E.~A.}\ \bibnamefont
  {Spiegel}}\ and\ \bibinfo {author} {\bibfnamefont {S.}~\bibnamefont
  {Zaleski}},\ }\href@noop {} {\bibfield  {journal} {\bibinfo  {journal}
  {Physics Letters A},\ }\textbf {\bibinfo {volume} {106}},\ \bibinfo {pages}
  {335} (\bibinfo {year} {1984})}\BibitemShut {NoStop}%
\end{thebibliography}%

\end{document}